\documentclass{aa}  

\usepackage{graphicx}
\usepackage{txfonts}
\usepackage{color}

\begin{document} 

\title{Metallicity dependence of HMXB populations}

\author{V.~M. Douna\inst{1,2}\fnmsep\thanks{\email{vdouna@iafe.uba.ar}}
\and L.~J. Pellizza\inst{3}
\and I.~F. Mirabel\inst{1,4}
\and S.~E. Pedrosa\inst{1}
}

\institute{Instituto de Astronom\'{\i}a y F\'{\i}sica del Espacio (IAFE), UBA-CONICET, CC 67, Suc. 28, (C1428ZAA), Buenos Aires, Argentina.\\
\and Facultad de Cs. Exactas y Naturales (FCEN), UBA, Buenos Aires, Argentina.\\
\and Instituto Argentino de Radioastronom\'{\i}a (IAR), CCT-La Plata, CONICET, C.C.5., 1894, Villa Elisa, Argentina.\\
\and CEA-Saclay, IRFU/DSM/Service d’Astrophysique, 91191 Gif/Yvette, France.
}

   \date{Received; accepted}

 
  \abstract
{High-mass X-ray binaries (HMXBs) might have contributed a non-negligible fraction of the energy feedback to the interstellar and intergalactic media at high redshift, becoming important sources for the heating and ionization history of the Universe. However, the importance of this contribution depends on the hypothesized increase in the number of HMXBs formed in low-metallicity galaxies and in their luminosities.}
{In this work we test the aforementioned hypothesis, and quantify the metallicity dependence of HMXB population properties.}
{We compile from the literature a large set of data on the sizes and X-ray luminosities of HMXB populations in nearby galaxies with known metallicities and star formation rates. We use Bayesian inference to fit simple Monte Carlo models that describe the metallicity dependence of the size and luminosity of the HMXB populations.}
{We find that HMXBs are typically ten times more numerous per unit star formation rate in low-metallicity galaxies ($12 + \log ({\rm O}/{\rm H}) < 8$, namely $<20\%$ solar) than in solar-metallicity galaxies. The metallicity dependence of the luminosity of HMXBs is small compared to that of the population size. }
{Our results support the hypothesis that HMXBs are more numerous in low-metallicity galaxies, implying the need to investigate the feedback in the form of X-rays and energetic mass outflows of these high-energy sources during cosmic dawn.}


\maketitle

\section{Introduction}

It has been proposed that the energy feedback of high-mass X-ray binaries (HMXBs) plays a major role in the thermal and ionization histories of the Universe \citep{power2009,mirabel2011,jeon2013,power2013,knevitt2014}, and that it significantly affects the cosmic star formation rate and the evolution of low-mass galaxies \citep{justham2012,artale2015}. The importance of these sources is based on the hypothesis that they are more numerous, and possibly more luminous, in low-metallicity stellar populations. This, coupled with the chemical evolution of the Universe, suggests that in the early Universe the contribution of HMXBs to the energy feedback into the interstellar and intergalactic media (ISM and IGM) would have been at least as large as that of other high-energy feedback sources, such as supernovae. 

However, direct evidence for the increase in the number and luminosity of HMXBs in low-metallicity stellar populations is still scarce. \citet{grimm2003} and \citet{mineo2012} have carried out exhaustive observational studies, compiling statistically significant samples of HMXBs in local galaxies. They found that the total X-ray luminosity of a HMXB population ($L_{\rm X}$) scales with the star formation rate (SFR) of its host galaxy, which at first sight seems to imply that there are no metallicity effects. However, their sample includes mainly high-SFR galaxies that, owing to the correlation between SFR and metallicity, would be high-metallicity systems. Hence, no metallicity effects are expected. Trying to overcome this limitation, \citet{kaaret2011} and \citet{brorby2014} studied a sample of local low-SFR, low-metallicity blue compact dwarfs (BCDs). They found that these galaxies host HMXB populations ten times more numerous per unit SFR, than galaxies in the sample of \citet{mineo2012}.

On the other hand, \citet{basu-zych2012} have shown that the $L_{\rm X}$--SFR relation evolves with redshift, which could be due to metallicity effects. \citet{fragos2013} used population synthesis models coupled to large scale cosmological simulations to investigate the emission of HMXB populations at different redshifts. They found that the integrated X-ray luminosity per unit SFR increases with redshift, exceeding that of other X-ray sources (low-mass X-ray binaries, active galactic nuclei) at high redshifts. In addition, \citet{basu-zych2013} showed that the relation between the ratio $L_{\rm X} / {\rm SFR}$ and the gas-phase metallicity of a sample of $z < 0.1$ Lyman-break analogs with slightly subsolar metallicities ($12 + \log({\rm O}/{\rm H}) \gtrsim 8.1$) is consistent with the theoretical expectations of \citet{fragos2013}.

Although previous works showed the existence of a metallicity dependence of HMXB populations, a discrimination of its effects on the size of these populations from those on the intrinsic luminosities is still lacking. To explore this issue, in this work we compile from the literature a large sample of local galaxies for which individual HMXBs can be resolved. Our sample extends those of previous works \citep{mineo2012,brorby2014} by including homogeneous estimations of the metallicities of the galaxies, hence allowing the study of a larger metallicity range. Moreover, our analysis extends the work of previous authors by requiring our models to match both the data of the HMXB population sizes and luminosities. In this paper we present the construction of the observational sample (Sect.~\ref{sample}), and investigate the dependence on metallicity of the properties of HMXB populations (Sect.~\ref{analysis}). We discuss the results and present our conclusions in Sect.~\ref{conclusions}.

\section{The sample}
\label{sample}

In order to investigate the dependence of the properties of HMXB populations on the metallicity of their progenitors, for 49 galaxies in the Local Universe we compiled from the literature the number of binaries, the integrated X-ray luminosities, star formation rates (SFR) and metallicities. Following \citet{mineo2012}, we selected galaxies that satisfy two criteria: a) high specific star formation rate (${\rm sSFR} = {\rm SFR} / M_*$), and b) small distance. As the production of HMXBs depends on the SFR of the galaxy \citep{grimm2003}, and the number of low-mass X-ray binaries is proportional to its stellar mass $M_*$ \citep{gilfanov2004}, the first criterion is useful to select galaxies in which HMXBs dominate the population of X-ray binaries. In particular, we chose ${\rm sSFR} > 10^{-10}\, {\rm yr}^{-1}$. The distance criterion was imposed because we are interested in galaxies where individual HMXBs can be resolved. The galaxies in our sample were chosen by requiring that their distance is $D < 65 \, {\rm Mpc}$, or their redshift is  $z < 0.015$.

The main source of objects for our sample is the work of \citet{mineo2012}. This work compiles the number of sources, luminosities (both of individual XRBs and the total integrated luminosity of the galaxy), and star formation rates for 29 nearby ($D < 40 \, {\rm Mpc}$), star-forming galaxies with ${\rm SFR} \sim 0.1-20\, M_\odot$. The sample of these authors is based on a homogeneous set of observations in X-rays, infrared and UV from \textit{Chandra}, Spitzer, GALEX and 2MASS, with very specific selection criteria.  We extended their sample in two ways. First, we compiled from the literature the metallicities of the galaxies when available. This left us with only 19 galaxies, most of them being high-metallicity systems. As a second step we added to this sample a set of low-metallicity galaxies because the investigation of a large abundance range is needed to assess clearly the effects of metallicity on HMXB populations. The low-metallicity galaxies come mainly from the work of \citet{brorby2014}, who 
studied a set of 25 nearby ($D < 30\, {\rm Mpc}$) low-metallicity BCDs. Another 5 galaxies from different works \citep{ghigo1983,thuan2004,winter2006,kaaret2011} were also added. We note that 20 of the 49 galaxies in our sample have no HMXBs detected; however these data still provide useful constraints on the metallicity dependence, as we will show in Appendix~\ref{app:model}. The compilation of abundances, together with the extension of the metallicity range and the use of both the observed population sizes and integrated luminosities, would allow us to go beyond the results of previous works. As the low- and high-metallicity samples were taken from different sources, with different uncertainties and selection effects, a standardization of the data is needed to make statistical comparisons and draw meaningful conclusions. In the following sections we describe the steps followed to make the data homogeneous, and the caveats involved in the process.

\subsection{Number of binaries and their luminosity}

As all the works from which we construct our sample give the data of individual sources in the galaxy, we took directly the observed number of sources $N$ and their individual X-ray luminosities from the aforementioned works (the complete data of our sample is given in Table~\ref{tabla}). The luminosity threshold $L_{\rm th}$ above which sources are detected was also recorded. Although we cannot ensure that all the sources listed are indeed HMXBs, the selection criteria imposed minimize the contamination by other types of sources. In the case of the data taken from \citet{mineo2012}, we only used objects with location flag equal to 1 in their Table~A1, which are those sources lying in the HMXB-dominated regions of the galaxies. For the data of low-metallicity galaxies, the association of the few X-ray sources found with star-forming regions makes highly improbable of them being other kind of sources.

\begin{table*}
\centering
\begin{tabular}{lccccccccc}
\hline\hline
Galaxy &   $D$   & $12 + \log({\rm O/H})$ &             ${\rm SFR}$            & $\log L^{\rm G}_{\rm X}$          &    $\log L_{\rm th}$   & $N$ & $\log L^{\rm G,36}_{\rm X}$ &  $\log L^{\rm G,38}_{\rm X}$ & $N^{38}$\\
 & [Mpc] &                       & $[M_\odot \text{yr}^{-1}]$ & $[{\rm erg\, s}^{-1}]$ & $[{\rm erg\, s}^{-1}]$ &                    &  $[{\rm erg\, s}^{-1}]$   & $[{\rm erg\, s}^{-1}]$      &  \\
 \hline\hline
DDO\,68 \tablefootmark{i} & 5.9 & 7.15  & 0.020 & 38.04 & 37.5 & 2 & 38.06 & - & 0\\
IZw\,18 \tablefootmark{i} & 18.2 & 7.18 & 0.068 & 39.58 & 37.87 & 1 & 39.58 & 39.58 & 1\\
SBS\,0335-052 & 54.3\tablefootmark{1} & 7.29\tablefootmark{a} & 0.44\tablefootmark{*} & 39.52\tablefootmark{1} & 38.95 \tablefootmark{1} & 1 & 39.57 & 39.55 & 2\\
SBS\,1129+576\tablefootmark{i} & 26.3 & 7.41 & 0.063 & 39.01 & 38.62 & 1 & 39.02 & 39.01 & 1\\
SBS\,0940+544\tablefootmark{i} & 22.1 & 7.48 & 0.022 & 38.99 & 38.45 & 1 & 39.00 & 38.99 & 1\\
RC2A\,1116+51\tablefootmark{i} & 20.8 & 7.51 & 0.040 & 39.51 & 38.52 & 1 & 39.51 & 39.51 & 1\\
HS1442+4250\tablefootmark{i} & 8.67 & 7.64 & 0.015 & 38.34 & 38.11 & 1 & 38.35 & 38.34 & 1\\
RC2A\,1228+12\tablefootmark{i} & 21.2 & 7.64 & 0.031 & 38.53 & 38.53 & 1 & 38.55 & 38.54 & 1\\
VIIZw\,403\tablefootmark{i} & 3.87 & 7.69 & 0.014 & 38.62 & 37.14 & 2 & 38.62 & 38.60 & 1\\
Mrk\,71 & 3.4\tablefootmark{2} & 7.85\tablefootmark{b} & 0.041\tablefootmark{**} & 38.34 \tablefootmark{2} & 38.3 \tablefootmark{2}  & 1 & 38.37 & 38.35 & 1\\
NGC\,4631\tablefootmark{ii} & 7.6 & 8.13\tablefootmark{c} & 4 & 39.71 & 37.09 & 23 & 39.74 & 39.65 & 6\\
NGC\,1569\tablefootmark{ii} & 1.9 & 8.19\tablefootmark{d} & 0.078 & 37.95 & 35.71 & 10 & 37.93 & - & 0\\
NGC\,1313\tablefootmark{ii} & 4.1 & 8.23\tablefootmark{e} & 0.44 & 39.69 & 36.82 & 8 & 39.69 & 39.68 & 4\\
NGC\,5253\tablefootmark{ii} & 4.1 & 8.24\tablefootmark{f} & 0.38 & 38.60 & 36.07 & 17 & 38.60 & 38.31 & 1\\
NGC\,4625\tablefootmark{ii} & 8.2 & 8.27\tablefootmark{c} & 0.09 & 38.01 & 37.01 & 4 & 38.04 & - & 0\\
NGC\,2403\tablefootmark{ii} & 3.1 & 8.34\tablefootmark{g} & 0.52 & 39.41 & 36.16 & 42 & 39.41 & 39.30 & 2\\
NGC\,4214\tablefootmark{ii} & 2.5 & 8.36\tablefootmark{h} & 0.17 & 38.41 & 36.19 & 14 & 38.42 & 38.17 & 1\\
NGC\,4490\tablefootmark{ii} & 7.8 & 8.36\tablefootmark{i} & 1.8 & 40.06 & 37.1 & 32 & 40.07 & 40.03 & 11\\
NGC\,3034\tablefootmark{ii} & 3.9 & 8.36\tablefootmark{c} & 10.5 & 39.86 & 36.86 & 54 & 39.90 & 39.75 & 12\\ 
NGC\,4038-39\tablefootmark{ii} & 13.8 & 8.4\tablefootmark{j} & 5.4 & 40.25 & 36.92 & 83 & 40.26 & 40.20 & 20\\
NGC\,7793\tablefootmark{ii} & 4 & 8.4\tablefootmark{k} & 0.29 & 38.36 & 36.55 & 9 & 38.38 & 38.20 & 1\\
NGC\,3310\tablefootmark{ii} & 19.8 & 8.44\tablefootmark{l} & 7.1 & 40.77 & 37.84 & 23 & 40.79 & 40.77 & 22\\
UGC\,5720\tablefootmark{ii} & 24.9 & 8.4\tablefootmark{m} & 1.8 & 39.57 & 38.36 & 4 & 39.66 & 39.60 & 6\\
NGC\,5457\tablefootmark{ii} & 6.7 & 8.55\tablefootmark{n} & 1.5 & 39.60 & 36.36 & 96 & 39.61 & 39.49 & 11\\
NGC\,3079\tablefootmark{ii} & 18.2 & 8.57\tablefootmark{h} & 6 & 39.65 & 37.98 & 14 & 39.80 & 39.65 & 14\\
NGC\,5194\tablefootmark{ii} & 7.6 & 8.58\tablefootmark{o} & 3.7 & 40.01 & 37.05 & 69 & 40.02 & 39.91 & 16\\
NGC\,5775\tablefootmark{ii} & 26.7 & 8.64\tablefootmark{p} & 5.3 & 40.52 & 38.01 & 25 & 40.54 & 40.52 & 25 \\
NGC\,4194\tablefootmark{ii} & 39.1 & 8.67\tablefootmark{h} & 16.8 & 40.42 & 38.64 & 4 & 40.56 & 40.49 & 29\\
NGC\,1672\tablefootmark{ii} & 16.8 & 8.97\tablefootmark{q} & 12 & 40.45 & 37.74 & 25 & 40.49 & 40.45 & 22\\
\hline
J\,081239.52+483645.3\tablefootmark{i} & 9.04 & 7.16 & 0.002 & -  &  38.21 & 0 & - & - & 0\\
UGC\,772\tablefootmark{i} & 11.5 & 7.24 & 0.012 & - &  38.38 & 0 & - & - & 0\\
J\,210455.31-003522.2\tablefootmark{i} & 13.7 & 7.26 & 0.007 & - &   38.56 & 0 & - & - & 0\\
UGCA\,292\tablefootmark{i} & 3.5 & 7.27 & 0.002 & - &  37.34 & 0 & - & - & 0\\
J\,141454.13-020822.9\tablefootmark{i} & 24.6 & 7.32 & 0.011 & - & 38.53 & 0 & - & - & 0\\
6dFJ\,0405204-364859\tablefootmark{i} & 11 & 7.34 & 0.013 & - &  38.33 & 0 & - & - & 0\\
HS\,0822+3542\tablefootmark{i} & 12.7 & 7.35 & 0.004 & - &  38.48 & 0 & - & - & 0\\
J\,085946.92+392305.6\tablefootmark{i} & 10.9 & 7.45 & 0.002 & - &  38.36 & 0 & - & - & 0\\
KUG\,0937+298\tablefootmark{i} & 11.2 & 7.45 & 0.015 & -  &  38.36 & 0 & - & - & 0\\
UGC\,4483\tablefootmark{i} & 3.44 & 7.54 & 0.004 & -  &  37.55 & 0 & - & - & 0\\
J\,120122.3+021108.5\tablefootmark{i} & 18.4 & 7.55 & 0.008 & -  & 38.58 & 0 & - & - & 0\\
KUG\,0201-103\tablefootmark{i} & 22.7 & 7.56 & 0.015 & - &  38.54 & 0 & - & - & 0\\
KUG\,1013+381\tablefootmark{i} & 19.6 & 7.58 & 0.073 & - &  38.56 & 0 & - & - & 0\\
SBS\,1415+437\tablefootmark{i} & 13.7 & 7.59 & 0.040 & - &  38.52 & 0 & - & - & 0\\
SBS\,1102+606\tablefootmark{i} & 19.9 & 7.64 & 0.035 & - &  38.54 & 0 & - & - & 0\\
KUG\,0942+551\tablefootmark{i} & 24.4 & 7.66 & 0.019 & - &  38.52 & 0 & - & - & 0\\
KUG\,0743+513\tablefootmark{i} & 8.6 & 7.68 & 0.032 & - &  38.14 & 0 & - & - & 0\\
Mrk\,209\tablefootmark{iii} & 5.7 & 7.77\tablefootmark{a} & 0.025 & -  & 37.96 & 0 & - & - & 0\\
UM\,461\tablefootmark{iii} & 13.4 & 7.85\tablefootmark{r} & 0.010 & -  & 37.66 & 0 & - & - & 0\\
IIZw\,40 & 14\tablefootmark{3} & 8\tablefootmark{b} & 1.32\tablefootmark{***} & -  & 39.86\tablefootmark{3} & 0 & - & - & 3\\

\hline
\end{tabular}
\tablefoot{\textsl{References:}
\tablefoottext{i}{\citet{brorby2014},}
\tablefoottext{ii}{\citet{mineo2012},}
\tablefoottext{iii}{\citet{kaaret2011},}
\tablefoottext{a}{\citet{izotov1999},}
\tablefoottext{b}{\citet{izotov2011},}
\tablefoottext{c}{\citet{calzetti2007},}
\tablefoottext{d}{\citet{kobulnickyskillman1997},}
\tablefoottext{f}{\citet{lopezsanchezesteban2010},}
\tablefoottext{g}{\citet{esteban2009},}
\tablefoottext{h}{\citet{eng2008},}
\tablefoottext{i}{\citet{pilyuginthuan2007},}
\tablefoottext{j}{\citet{mirabel1992},}
\tablefoottext{k}{\citet{bibbycrowther2010},}
\tablefoottext{l}{\citet{denicolo2002},}
\tablefoottext{m}{\citet{hirashita2002},}
\tablefoottext{n}{\citet{cedres2004},}
\tablefoottext{o}{\citet{bresolin2004},}
\tablefoottext{p}{\citet{werk2011},}
\tablefoottext{q}{\citet{storchi1996},}
\tablefoottext{r}{\citet{kniazev2004},}
\tablefoottext{1}{\citet{thuan2004},}
\tablefoottext{2}{\citet{winter2006},}
\tablefoottext{3}{\citet{ghigo1983},}
\tablefoottext{*}{\citet{prestwich2013},}
\tablefoottext{**}{\citet{hopkins2002},}
\tablefoottext{***}{\citet{sage1992}.}}
\caption{Distances, oxygen abundances, SFRs, X ray luminosities, and observed number of HMXBs for the  galaxies in our sample.}
\label{tabla}
\end{table*}

X-ray luminosities of individual sources are given in different energy ranges in the literature, according to the instrument used to observe them. To make the data homogeneous, we corrected all the observations to the $0.5-8\, {\rm keV}$ band, which is that of most observations of our sample. To perform the correction, we assumed that the X-ray spectrum of individual sources is a power-law, and used for each source the spectral index reported in the literature. In the case it is not available, we took the standard value $-1.7$ \citep{fabbiano2006}. The luminosity thresholds of the observations were corrected accordingly. Throughout this paper, we will express all the luminosities in the $0.5-8\, {\rm keV}$ band.

The total X-ray luminosity $L^{\rm G}_{\rm X}$ of each galaxy in our sample is taken as the sum of the luminosities $L_i$ of individual sources above $L_{\rm th}$. However, the latter differs for each galaxy. Aiming at obtaining a homogeneous luminosity sample and following \citet{mineo2012}, we also computed the total luminosities of the galaxies corrected to a common threshold $L_{\rm th,1} = 10^{36}\, {\rm erg\, s}^{-1}$. In the case of NGC~1569, as $L_{\rm th} < L_{\rm th,1}$, this correction was made by subtracting the luminosity of the sources below $L_{\rm th,1}$. For the remaining galaxies, we used a power-law X-ray luminosity function (XLF) with the index ($\alpha = -1.58$) and normalization factor ($q = 1.49 \, (10^{38} \,{\rm erg \, s}^{-1})^{0.58} M_\odot^{-1} \, {\rm yr}$) given by \citet{mineo2012}, hence obtaining the corrected luminosity

\begin{equation}
\label{ecSens}
L^{\rm G,36}_{\rm X} = L_{\rm X}^{\rm G} + q \, {\rm SFR} \, \int_{L_{\rm th, 1}}^{L_{\rm th}} L^{\alpha + 1} \, dL.
\end{equation}

\noindent
In most cases, the integral in the previous equation amounts to less than 11\% of the final value, except for UGC~5720 (19\%), NGC~4194 (28\%), and NGC~3079 (29\%). We conservatively assumed here that the XLF does not depend on metallicity. This choice is not crucial because the corrections are small, owing to the weak dependence of the integrated luminosity on $L_{\rm th}$. However, if the higher normalization factor given by \citet{brorby2014} were used for low-metallicity galaxies, a slightly larger total luminosity would have been found, making our conclusions stronger.

It should be pointed out that the aforementioned correction on the luminosities was not applied to the galaxies without detected sources. In this case, any trend found in the corrected values would reflect the corresponding trends of the model used in Eqn.~\ref{ecSens} for the XLF, rather than those of the original data, hence biasing the results.

Following the above reasoning, we also corrected the number of sources detected in each galaxy to a common threshold $L_{\rm th,2} = 10^{38}\, {\rm erg\, s}^{-1}$. The choice of this threshold value was dictated by the fact that the observed number of sources does depend strongly on $L_{\rm th}$. Hence, we took a large value so that for most galaxies $L_{\rm th} \leq L_{\rm th,2}$, and the number of sources $N^{38}$ above $L_{\rm th,2}$ is determined by simple counting. Only for 26 of our 49 galaxies is $L_{\rm th} > L_{\rm th,2}$ and then a correction based on the XLF is needed,

\begin{equation}
\label{ecSensnum} 
N^{38} = N + q \, {\rm SFR} \, \int_{L_{\rm th, 2}}^{L_{\rm th}} L^{\alpha} \, dL.
\end{equation}

\noindent
For 22 of them, the integral in the right-hand side of Eqn.~\ref{ecSensnum} is $\ll 1$, and $N^{38} = N$. In the few remaining cases, the corrections amount to 33\% (UGC~5720), 50\% (SBS~0335-052), 86\% (NGC~4194), and 100\% (IIZw~40) of the final value. The luminosity $L^{\rm G,38}_{\rm X}$ was determined in the same way as $L^{\rm G,36}_{\rm X}$, but using $L_{\rm th,2}$ instead of $L_{\rm th,1}$ in Eqn.~\ref{ecSens}. The corrections in this case are lower than 7\%, except for NGC~4194 (15\%).

The corrections made by Eqns.~\ref{ecSens} and \ref{ecSensnum} allow us to obtain a uniform sample of the numbers of sources and luminosities of the galaxies. It is important to note that most of these corrections are small, and that they do not affect our final results, as $N^{38}$, $L^{\rm G,38}_{\rm X}$, and $L^{\rm G,36}_{\rm X}$ are used only to visualize and derive preliminary trends. The full Bayesian analysis of the data, from which our final results are obtained and our conclusions are drawn, relies on the original data ($N$, $L^{\rm G}_{\rm X}$, and $L_{\rm th}$).

\subsection{Star formation rates}

We extracted the star formation rates (SFRs) of 19 of the galaxies of our sample from \citet{mineo2012}. These authors use a combined estimator for the SFR based on the near-UV (NUV) radiation from young stars, corrected by the IR emission due to heating of dust,

\begin{equation}
\rm{SFR} = {\rm SFR_{NUV}} + (1 - \eta) {\rm SFR_{IR}},
\end{equation}

\noindent
where $\eta$ is a correction factor related to the IR emission coming from the old stellar population, whose value is null for starburst galaxies and 0.4 for normal disk galaxies. The NUV and IR contributions to SFR (in $M_\odot \, {\rm yr}^{-1}$) can be calculated from the $2312{\AA}$ and $8 - 1000 \, \mu{\rm m}$ luminosities (in ${\rm erg\, s}^{-1}$) as ${\rm SFR}_{\rm IR} = 4.6 \times 10^{-44} L_{\rm IR}$ and ${\rm SFR}_{\rm NUV} = 1.2 \times 10^{-43} L_{\rm NUV}$, respectively. In addition to the galaxies taken from the work of \citet{mineo2012}, we adopted this estimator for those galaxies for which NUV and IR observations are available. For the rest of the galaxies we used far-UV (FUV) ${\rm SFR}_{\rm FUV}$ estimations corrected by a factor of 1.23. According to \citet{prestwich2013} and \citet{brorby2014}, this factor gives the ratio of FUV estimations to those based on NUV and IR observations. For a single galaxy (II~Zw~40) the only available ${\rm SFR}$ estimation is based on ${\rm H}_\alpha$ measurements. In 
this case, we transformed this value into a FUV ${\rm SFR}$ using the conversion of \citet{hunter2010}, and then once again using the factor 1.23 to transform it to the NUV/IR ${\rm SFR}$ scale. In this way, we obtained a sample as homogeneous as possible to compare the effects of SFR in the production of HMXBs.

\subsection{Metallicities}

The metallicity estimations for our sample were taken directly from the literature. As we are dealing with star-forming galaxies, in which emission lines are used to estimate abundances, we use the oxygen abundances $12 + \log ({\rm O}/{\rm H})$  as a metallicity indicator. The abundance values compiled are {\em a priori} far from homogeneous, as they were obtained with different methods whose systematic differences are still unclear \citep{bresolin2004, garnett2004, liang2006, yin2007a, yin2007b, kewleyellison2008, lopezsanchez2012}. Although it is still not certain which method is better to estimate the absolute oxygen abundance of a galaxy, it is believed that the relative metallicities between galaxies estimated with the same calibration are reliable \citep{kewleyellison2008}. However, due to the large metallicity range of the galaxies in our sample, it was not possible to calculate the oxygen abundances with a single method because of the lack of homogeneity of the spectral lines observed. 

It is important to note that in this work we are not interested in the computation of absolute oxygen abundances, but in the correlation of metallicity with the X ray emission of galaxies. With this aim and in order to evaluate the impact of comparing abundances estimated by different calibrations, we searched the literature for the raw spectral information of each galaxy, and built a homogeneous subsample of abundances computed by a single method. To this aim we used the intensities of the spectral lines ${\rm [NII]}\,\lambda\,6583$ and ${\rm H}_\alpha$, and computed the abundances of 21 galaxies using the ${\rm N}_2$ index \citep{pettinipagel2004}. This method is single-valued, and is based on a ratio between lines that are very near in the spectrum, hence it does not depend on the reddening correction or flux calibration. 

\begin{figure}
\centering
\includegraphics[width=\hsize]{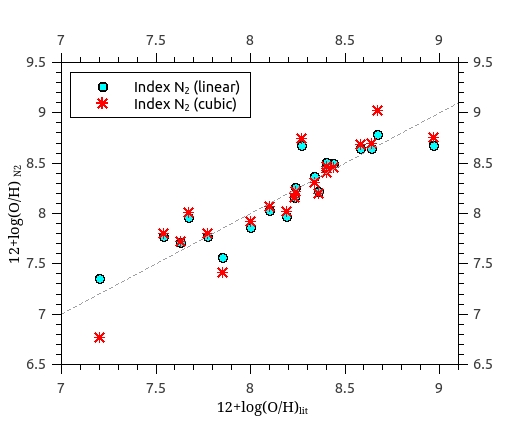}
\caption{Oxygen abundances calculated from raw spectral data by using both fits to the ${\rm N}_2$ ratio given by \citet{pettinipagel2004}, as a function of the abundances taken directly from the literature. Cyan dots (red stars) correspond to the linear (cubic) calibration of the ${\rm N}_2$ index; the dashed line represents the identity.}
\label{MetN2}
\end{figure}

For those 21 galaxies of our homogeneous subsample, we calculated the abundances both using the linear and cubic fits for the ${\rm N}_2$ ratio given by \citet{pettinipagel2004}. The correlation between both ${\rm N}_2$ estimations and the abundances taken directly from the literature are shown in Fig.~\ref{MetN2}. The variance between them is $0.17 \, {\rm dex}$ in the case of the linear method and $0.22 \, {\rm dex}$ for the cubic fit. Hence, despite the possible systematic deviations between the different calibrations, the general trend observed supports the use of the complete sample to estimate relative metallicities.  

\section{Analysis of the sample}
\label{analysis}

\subsection{Observed trends}
\label{sec:trends}

\begin{figure}
\centering
\includegraphics[width=\hsize]{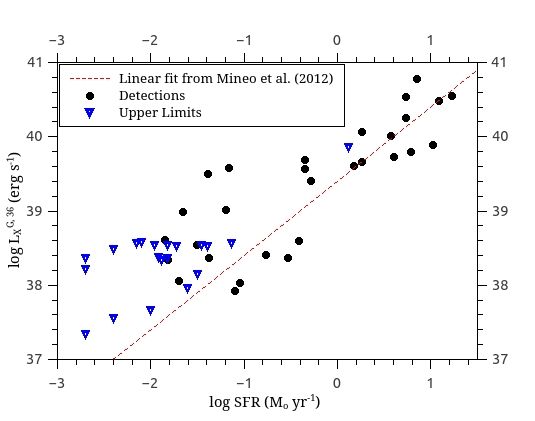}
\caption{X-ray luminosity as a function of SFR for the sample of galaxies compiled in this work. Black circles correspond to those galaxies with measured X-ray luminosities, while blue triangles represent the detection limits for single sources. These can be regarded as upper limits for the galaxy luminosity in the case it had only one source. The red dashed line ($\log L_{\rm X} = \log {\rm SFR} + 39.4$) is the relation found by \citet{mineo2012}. It describes well the data at high SFRs, whereas it fails to do so in the opposite SFR range. For $\log {\rm SFR} \lesssim -1.1$, all the data fall above the fit, which suggests the existence of metallicity effects in the sample.}
\label{lxvssfr}
\end{figure}

In Fig.~\ref{lxvssfr} we show the $L_{\rm X}$--SFR relation for the galaxies in our sample, together with the best fit of \citet[$\log L_{\rm X} = \log {\rm SFR} + 39.4$]{mineo2012}. In order to make a meaningful comparison, in this plot we use $L^{\rm G,36}_{\rm X}$ so that all galaxies have the same luminosity threshold, which is also the one used by these authors. It is evident that their best fit describes properly the mean behavior of high-SFR galaxies, mostly taken from their work, but fails to describe the correlation at low SFRs. In this limit, the X-ray luminosities are higher than those predicted by the fit. Indeed, all galaxies with $\log {\rm SFR} \lesssim -1.1$ fall above the fit of \citet{mineo2012}. The fit also fails to describe the large dispersion of the data in the whole metallicity range. As the SFR of galaxies is correlated to their mean metallicity, the increase in the X-ray luminosity with respect to the fit of \citet{mineo2012} observed at low SFRs could be due to metallicity effects.
 
However, it could also be due to statistical fluctuations, as low-SFR galaxies have small HMXB populations. The increase in the dispersion at low SFRs suggests that this effect is present in our sample. Disentangling both effects is crucial to reach a meaningful conclusion about the metallicity dependence of HMXB populations.

\begin{figure}
\centering
\includegraphics[width=\hsize]{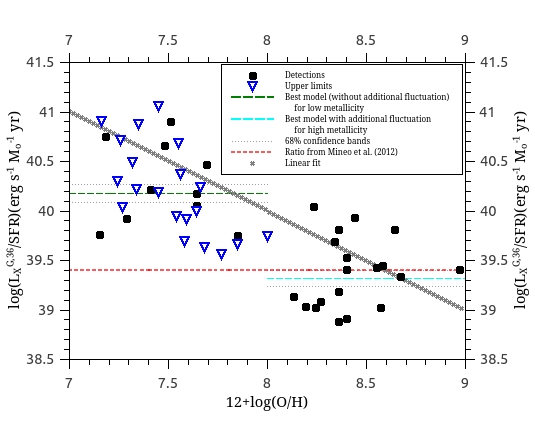}
\caption{X-ray luminosity per unit SFR as a function of the oxygen abundance of the galaxies in our sample. Black filled squares correspond to those galaxies with measured X-ray luminosities, while blue triangles represent the detection limits for single sources. These can be regarded as upper limits for the galaxy luminosity in the case it had only one source. The red short-dashed line represents the value found by \citet{mineo2012} for solar-metallicity galaxies ($L^{\rm G,36}_{\rm X} / {\rm SFR} = 39.4$), while the gray dotted line is the best linear fit to our sample, $\log L^{\rm G,36}_{\rm X} = 39.26 + \log {\rm SFR} - 1.01 [\log ({\rm O}/{\rm H}) - \log ({\rm O}/{\rm H})_\odot]$. A clear trend of decreasing $L^{\rm G,36}_{\rm X} / {\rm SFR}$ with oxygen abundance is observable, which is a direct evidence of the existence of metallicity effects in HMXB populations. Stochasticity effects are also evident, and require a more refined treatment to make a robust prediction of the variation of the $L^{\rm G,36}_{\rm X} / {\rm SFR}$ ratio with metallicity. The cyan and green long-dashed lines above and below $12+\log ({\rm O}/{\rm H}) = 8$ respectively, represent our best models for the high- and low-metallicity subsamples, with the 68\% confidence bands for the parameter $Q$ (small-dotted gray lines).}
\label{lxsfrvsmet}
\end{figure}

To further investigate the influence of metallicity on the $L_{\rm X}$--SFR relation, in Fig.~\ref{lxsfrvsmet} we show the dependence of the $L^{\rm G,36}_{\rm X} / {\rm SFR}$ ratio on the oxygen abundance of the galaxies. Despite the high dispersion, the observed behavior is consistent with an anticorrelation between the X-ray luminosity per unit SFR and the oxygen abundance. It is also apparent that a linear relation is a good description of the correlation. The best linear fit (not including upper limits) is $\log L^{\rm G,36}_{\rm X} = 39.26 + \log {\rm SFR} - 1.01 [\log ({\rm O}/{\rm H}) - \log ({\rm O}/{\rm H})_\odot ]$, taking $12 + \log ({\rm O}/{\rm H})_\odot = 8.69$ from \citet{asplund2009}. The $L^{\rm G,36}_{\rm X} / {\rm SFR}$ ratio is similar to that of \citet{mineo2012} at solar metallicity, while it increases about $1 \, {\rm dex}$ per dex of decrease in abundance. However, the dispersion around the regression line is still important ($\sim 0.5\,{\rm dex}$), suggesting that stochasticity 
effects should be investigated to make a more robust prediction of the variation with metallicity of the X-ray luminosity per unit SFR of galaxies.

\begin{figure}
\centering
\includegraphics[width=\hsize]{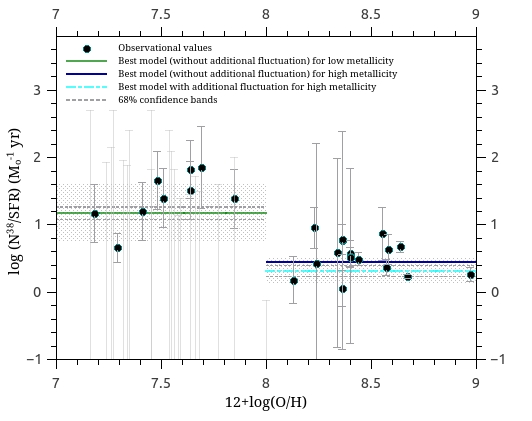}
\caption{Observed number of HMXBs with luminosities higher than $10^{38} \, {\rm erg \, s}^{-1}$, per unit SFR, as a function of the oxygen abundance. Black circles are the data points, with error bars representing Poissonian uncertainties. Large error bars reaching the lower end of the plot represent galaxies with no detected sources; their upper ends correspond to $N^{38} = 1$. Our best model for low-metallicity galaxies ($12 + \log ({\rm O}/{\rm H}) < 8$, green solid line, $Q = 3.13$, $\sigma_{\rm extra} = 0$) predicts ten times more HMXBs than the corresponding best model for high-metallicity galaxies (cyan dot-dashed line, $Q = 2.14$, $\sigma_{\rm extra} = 22$). For both models, the 68\% confidence bands for the parameter $Q$ and the dispersion predicted for the data are shown (gray dashed lines and shaded regions, respectively). The mean $Q=2.28$ for high-metallicity galaxies with $\sigma_{\rm extra} = 0$ is also plotted (blue solid line). The difference between low- and high-metallicity galaxies cannot be explained by statistical effects.}
\label{n-sfrvsmet}
\end{figure}

The increase in the $L_{\rm X} / {\rm SFR}$ value toward low oxygen abundances might be due to an increase either in the HMXB population size, or in the mean HMXB luminosity. To disentangle these two physical effects, in Fig.~\ref{n-sfrvsmet} we show the observed number of sources $N^{38}$ with luminosities above the fixed threshold $L_{\rm th,2} = 10^{38}\, {\rm erg\, s}^{-1}$, per unit SFR, as a function of the metallicity of the galaxies. A clear trend is seen, in the sense that this number is lower for high-metallicity galaxies ($\log ({\rm O}/{\rm H}) > 8$) than for low-metallicity ones ($\log ({\rm O}/{\rm H}) < 8$). The difference is about $1\,{\rm dex}$. The presence of many data points with error bars much smaller than this difference suggests that it cannot be explained by statistical effects. As the observed number of sources depends mainly on the population size, and only weakly in the HMXB luminosities, this result suggests that the main metallicity effect is a change in the total number of HMXBs per unit SFR. Indeed, the observed difference in $N^{38}/{\rm SFR}$ from low- to high-metallicity galaxies is roughly the same amount as the change in $L^{\rm G,36}_{\rm X} / {\rm SFR}$ derived from Fig.~\ref{lxsfrvsmet}. This implies that the increase in the integrated luminosity of galaxies at low metallicities can be explained by the increase in the size of their HMXB populations. An increase in the mean luminosity of HMXBs, if any, is certainly much lower. Once again, the dispersion of the $N^{38}/{\rm SFR}$ data is high, indicating the presence of stochasticity effects which must be taken into account to obtain a more robust prediction of the amount of the observed variations with metallicity.

\subsection{Stochasticity effects}
\label{sec:stochasticity}

As pointed out before, all the data shown in this section present large dispersions. Part of them may be due to the stochasticity of the observed HMXB populations, which affects both their sizes and observed luminosities. Indeed, it has been shown by \citet{gilfanov2004b} that the collective luminosity of a small population of randomly chosen sources can grow non-linearly with the number of sources, contrary to the prediction of models based on infinite populations. To obtain a robust description of the trends present in the data, we developed a simple stochastic model for the number of sources and total luminosity of the population of HMXBs of a star-forming galaxy with a given SFR and oxygen abundance, which attempts to describe the observed behavior (see Appendix~\ref{app:model} for its details). Following \citet{brorby2014}, we assume that the mean number of sources in a galaxy follows a Poisson distribution whose mean scales with its SFR,

\begin{equation}
\label{Neq}
\langle N_{\rm HMXB} \rangle = Q \, {\rm SFR},
\end{equation}

\noindent
where $Q$ is a free parameter that may depend on the oxygen abundance. We also assume for the HMXBs a standard power-law XLF with index $\alpha = -1.6$ in the range $[L_{\rm min}, L_{\rm max}] = [10^{35}, 10^{40}] \, {\rm erg \, s}^{-1}$ \citep{fabbiano2006,mineo2012}. The observed number of sources and integrated luminosity of a particular galaxy above any specified threshold luminosity can be computed using the XLF. We note that our model includes only the dependence of the population size on metallicity, and not that of HMXB luminosities, as the XLF is fixed. In addition, our model includes two fluctuation sources, an intrinsic one due to the stochasticity of the star formation process, and another one given by the detection process, which selects only those binaries with luminosities above $L_{\rm th}$. Model fitting was done by a Markov-chain Monte Carlo method, which gives the best-fit value of $Q$, together with its variance. 

As \citet{brorby2014} proposed that $Q$ is different for low- and high-metallicity galaxies, we divide our sample into two subsamples comprising galaxies below and above $12 + \log ({\rm O}/{\rm H}) = 8$, respectively. We fit our model to each sample, obtaining $\log Q_{\rm low} = 3.13 \pm 0.13$ and $\log Q_{\rm high} = 2.28 \pm 0.02$ \citep[which correspond to $q_{\rm low} = 12.8$ and $q_{\rm high} = 1.81$ using the notation of][]{brorby2014}\footnote{The factor $Q$ in {\rm our} work differs from that used by \citet{brorby2014} in the normalization of the luminosity function.}. This results agree reasonably well with those of \citet{mineo2012} ($q = 1.49$) and \citet{brorby2014} ($q = 14.5$), and imply that low-metallicity galaxies produce $\sim 10$ times more HMXBs per unit SFR that their high-metallicity counterparts.

The dispersion given by our model describes the one of the low-metallicity subsample very well, as shown in Fig.~\ref{n-sfrvsmet}. However, this is not true for the high-metallicity data. To investigate this issue, we added an extra dispersion $\sigma_{\rm extra}$ to the number of sources predicted by our model, and fit it to the high-metallicity subsample for different (fixed) values of $\sigma_{\rm extra}$. A Bayesian model comparison shows that the data are best described by $\sigma_{\rm extra} = 22$, typically 3~times that given by Poissonian effects, and a slightly lower $\log Q_{\rm high} = 2.14$. The decrease in the latter value compared to the $\sigma_{\rm extra}=0$ model is due to the fact that the extra dispersion (constant and larger than the Poissonian uncertainties of the data) increases the weight of lower-$N$ values relative to higher-$N$ ones, hence lowering the best-fit value. The fact that an extra dispersion is not observed in the low-metallicity sample could be explained by the different sample sizes (556 high- vs. 12 low-metallicity sources). Indeed, the test of picking random samples of 12 high-metallicity sources and performing the same analysis shows that the extra dispersion remains undetected if the size of the high-metallicity sample is small. Moreover, the mean best-fit value of $Q_{\rm high}$ obtained for the random samples of 12 high-metallicity sources agrees to within uncertaintes with the best-fit value of the whole high-metallicity sample. This fact implies that the increase in the number of HMXBs per unit SFR at low metallicities is not an effect of the difference in the sample sizes.

The aforementioned results confirm that the metallicity dependence observed in our sample is not an effect of statistical fluctuations, but a real physical trend, as the value of $Q$ is different for the low- and high-metallicity subsamples even when the dispersion of the data is well described. Indeed, the probability that the low-metallicity data comes from random fluctuations around the best high-metallicity model (including the extra dispersion) is $\sim 10^{-47}$ times that given by the best low-metallicity model. Hence, in agreement with \citet{brorby2014}, our analysis suggests that the number of HMXBs per unit SFR increases by a factor of $\sim 10$ at low metallicities, even if we take into account the statistical fluctuations. As an interesting by-product, these fluctuations are well described by Poissonian effects at low metallicities but are larger at high metallicities. 

The models that best fit the number of sources of the galaxies also describe nicely their integrated luminosities (Fig.~\ref{lxsfrvsmet}). In the high metallicity limit, the predictions are also in well agreement with the relation of \citet{mineo2012}. This result suggests that the most numerous HMXB populations produced at low metallicities can explain not only the different observed number of sources, but also the enhanced integrated X-ray luminosity of galaxies. Hence, the dependence of the instrinsic luminosity of the sources with metallicity would be small. However, given that the {\em observed} number of HMXBs depends on the XLF through the detectability of the sources, and that the XLF is fixed in our model, a dependence of the luminosity on metallicity cannot be completely discarded.
 
To test the consistency of the observed luminosities with those predicted by our model in a statistical way, we extended our model to predict the luminosities of the observed HMXBs by sampling the XLF with the restriction $L > L_{\rm th}$ (see Appendix~\ref{app:model} for details). We fit the model to the high- and low-metallicity subsamples, now including the observed luminosities of the galaxies together with their numbers of HMXBs in the comparison. We included the previously obtained extra dispersion in the high-metallicity model. The resulting values of the free parameter $Q$ are in full agreement with the ones obtained before, confirming that the luminosity dependence on metallicity is small, if any.

\begin{figure}
\centering
\includegraphics[width=\hsize]{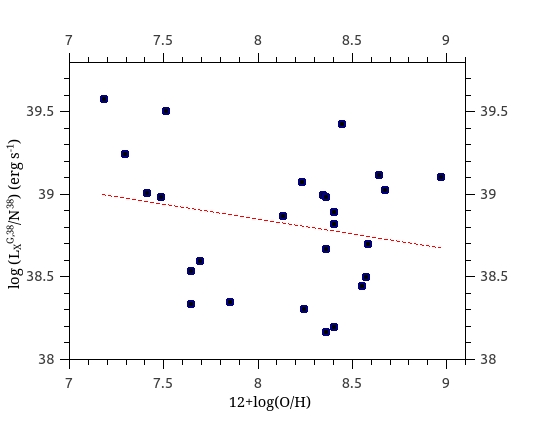}
\caption{Mean luminosity of HMXBs above a threshold of $10^{38} \, {\rm erg \, s}^{-1}$, as a function of the oxygen abundance. Black, filled squares are the data points, while the red dashed line is the best linear fit to them, $\log L_{\rm X}^{G,38} / N^{38} = (38.7 \pm 1.3) - (0.18 \pm 0.16) [\log ({\rm O}/{\rm H}) - \log ({\rm O}/{\rm H})_\odot]$. A low-significance anticorrelation between these variables is observed, which might suggest a brightening of HMXBs at low metallicities. As in previous figures, the dispersion around the fit is large ($0.4 \, {\rm dex}$).}
\label{l-nvsmet}
\end{figure}

To further explore this possibility, we analyzed the correlation between the observed mean luminosity of HMXBs in each galaxy ($L_{\rm X}^{\rm G, 38} / N^{38}$) and its metallicity (Fig.~\ref{l-nvsmet}). This figure shows a very weak trend (red, dashed line) in the sense that HMXBs are brighter in low-metallicity populations, masked by large-amplitude fluctuations. If this trend could be confirmed by more numerous and more precise data, then it might indicate that the observed variation of the total X-ray luminosity of the HMXB populations results from a large contibution due to the change in the population size, plus a minor one originated in the variation of the mean HMXB luminosity.

Finally, it is important to point out that although the simplest model that describes the dependence of the number of HMXBs on the oxygen abundance is a split function, there is no physical evidence to justify the break at $12 + \log ({\rm O} / {\rm H}) \simeq 8$. Hence, following the apparent linear trend shown by the number of observed HMXBs in Fig. ~\ref{n-sfrvsmet}, as a second approach we also parameterized the factor $Q$ by a linear function of the oxygen abundance. A fit to our whole sample confirms the existence of a dependence of $Q$ on metallicity with a confidence greater than $1 - 5 \times 10^{-5}$. However, the best-fit parameters suggest that the anticorrelation between the number of binaries and the metallicity of the galaxies is milder than that predicted by a sharp break. In this case the change in the population size per unit SFR is only $-0.14\,{\rm dex}$ per dex. The discrepancy is originated in the better statistics of high-metallicity data, which have more weight in the fit than low-metallicity ones, hence biasing the slope to shallower values. A deeper exploration of this issue would require larger samples, and a more detailed treatment of population size fluctuations.

\section{Conclusions}
\label{conclusions}

We compiled from the literature a large sample of 49 galaxies with measured star-formation rates, oxygen abundances and HMXB population properties (observed number of sources and luminosities above some threshold luminosity). This sample enhances those previously compiled by \cite{mineo2012} and \citet{brorby2014} by giving the oxygen abundances of the galaxies, and by including both the number of sources in each galaxy together with its integrated luminosity. As the low- and high-metallicity galaxies were taken from different works, with different uncertainties and selection effects, the data were homogenized to make statistical comparisons and draw meaningful conclusions. The SFRs in our sample were corrected to a common estimator. For the oxygen abundances, we used a single estimator in the largest possible subsample and the rest of the sample was shown to be unbiased with respect to this homogeneous set. Any residual non-uniformity that may have been left over by this process is small, and does not affect our conclusions. Hence, our complete sample is then suitable to investigate the effects of metallicity in the properties of HMXB populations.

Our main results are the following:

\begin{itemize}

\item
The size of the HMXB populations of low-metallicity ($12 + \log ({\rm O} / {\rm H}) < 8$) galaxies, per unit SFR, is $\sim 10$~times larger than that of solar metallicity galaxies. This result is robust; it describes well the observed number of HMXBs and integrated X-ray luminosities of nearby star-forming galaxies, taking into account the observed fluctuations. It does not depend on the different sizes of the low-and high-metallicity subsamples. It is also in excellent agreement with those of \citet{kaaret2011} and \citet{brorby2014}, who found similar results using slightly different methods and data.

\item
The dispersion in the observed number of HMXBs of low-metallicity, low-SFR galaxies can be explained by Poissonian fluctuations associated with the initial mass function and XLF sampling. This is not the case for high-metallicity galaxies, which show fluctuations typically half an order of magnitude higher than those expected from these effects. \citet{mineo2012} have reported an analog dispersion in the $L_{\rm X}$--SFR relation of the high-metallicity subsample, which cannot be attributed to selection effects. Our results indicate that the large dispersion is not an effect of the IMF or XLF sampling. Therefore it could be due to a different physical origin, most probably the internal metallicity dispersion of the galaxies and the age distribution of HMXBs. Indeed, numerical simulations of stellar populations which compute the effect of these fluctuations give dispersion values that agree with our findings (Artale et al., in prep.). This effect is hidden in the low-metallicity subsample behind the large Poissonian fluctuations that arise from the small number of sources.

\item
Present evidence suggests that the aforementioned metallicity effects occur below $12 + \log ({\rm O} / {\rm H}) \approx 8$. Although a smooth linear trend in the correlations is apparent, it is not supported by statistical tests on the present sample. A sharp break at this metallicity provides the best description of our data. However, the smoothness of the metallicity dependence could be affected by differences in the sample size or residual inhomogeneities.  A larger sample including galaxies near the break and more low-metallicity galaxies is crucial to solve this issue. A more thorough description of the data fluctuations, especially those from non-Poissonian sources, would also be desirable.

\item
The dependence of the luminosity of HMXBs with metallicity, if any, is much smaller than that of the population size. However, refined models are needed to determine its significance.

\end{itemize}

Our results support the general picture describing the enhancement of the HMXB population size at low metallicities \citep{dray2005, linden2010, kaaret2011,basu-zych2012,basu-zych2013,fragos2013,brorby2014}. This implies that these energetic sources were more numerous in the Early Universe, hence a promising source of heating and partial reionization of the IGM, as proposed by several authors \citep{power2009,mirabel2011,jeon2013,power2013,knevitt2014}. It also implies that the energy feedback from HMXBs to the ISM could play an important role in the formation and evolution of low-mass galaxies, supporting recent results \citep{artale2015}. To assess the exact magnitude of these effects, a more detailed description of the variation of the number of HMXBs and their luminosities must be given. This is only possible with further deep observations that unveil the low-luminosity end of the XLF in nearby galaxies, together with homogeneous metallicity and SFR estimations. Age measurements of the star-forming 
regions would also be valuable to investigate the 
effects of this important parameter.


\begin{acknowledgements}
We would like to thank the referee for useful comments that helped to clarify the manuscript. This work was partially supported by Argentine grants PICT~2011-0959 (ANPCyT), PIP~2009-0305 and PIP~2012-0396 (CONICET). Our MC models were run in the Fenix cluster of the Numerical Astrophysics Group at the Institute for Astronomy and Space Physics (CONICET/UBA).
\end{acknowledgements}

\appendix

\section{Monte Carlo model for HMXB populations}
\label{app:model}

\subsection{The model}

Our model aims at describing the observed number of sources and total luminosity of the population of HMXBs of a star-forming galaxy with a given SFR and oxygen abundance, to perform a robust comparison with observations. We start from the fact that the progenitors of HMXBs are massive stars, and that these sources have short evolutionary times \citep[less than $\sim 100 \, {\rm Myr}$;][]{belczynski2004, belczynski2008, belczynski2010, sty2005, sty2007}. Hence, their number should correlate with the SFR, as \citet{brorby2014} propose in their Eqn.~4. However, star formation can be regarded as a stochastic process in which the final result (the formation of a single or binary star, and in the last case the initial masses of the components, the semimajor axes and eccentricities, etc.) follows a certain probability distribution. The sampling of this distribution in a particular stellar population is a source of fluctuations in the number of HMXBs ($N_{\rm HMXB}$) of a galaxy. If we assume that the probability 
$p_{\rm HMXB}$ of producing a HMXB is fixed, then $N_{\rm HMXB}$ has a Bernoulli distribution. As the total number of stars $N_\star$ in any stellar population is large, and $p_{\rm HMXB} \ll 1$ because the initial mass function is a strongly decreasing function, we can approximate $N_{\rm HMXB}$ as a Poisson variable. The mean value of $N_{\rm HMXB}$ is then $\langle N_{\rm HMXB} \rangle = N_\star p_{\rm HMXB}$. However, as it is not our aim to compute the distribution of $N_{\rm HMXB}$ from first principles, following \citet{brorby2014} we parameterize it as

\begin{equation}
\label{Neq2}
\langle N_{\rm HMXB} \rangle = Q(\log({\rm O}/{\rm H})) \, {\rm SFR},
\end{equation}

\noindent
as $N_\star$ scales with SFR. The factor $Q$ is the same as $q$ in the Eqn.~4 of \citet{brorby2014}, apart from the normalization factor of the XLF. We note that the difference between our model and that of these authors is that the above equation decribes the mean number of sources in a Poisson process, while the corresponding equation of \citet{brorby2014} describes the exact number. When $N \sim 1$ (as in low-SFR galaxies) differences arise, and the stochasticity must be taken into account. Our factor $Q$ may depend continuously on the oxygen abundance, or just assumed to be different for low and high metallicities. 

A second source of stochasticity is given by the detection process, which selects only those binaries with luminosities above some threshold $L_{\rm th}$. The XLF sampling in the HMXB population introduces statistical fluctuations in the number of sources detected. Given that detection is a Bernoulli process, the composition with the Poisson process described above produces a new Poisson process with mean

\begin{equation}
\label{nobs}
\langle N_{\rm det} \rangle = p_{\rm det} Q(\log(O/H)) \, {\rm SFR},
\end{equation}

\noindent
where

\begin{equation}
\label{pdet}
p_{\rm det} = k_{\rm norm} \int_{L_{\rm th}}^{L_{\rm max}} \psi (L) dL.
\end{equation}

\noindent
Here the XLF is $\psi(L) = k^{-1}_{\rm norm} L^\alpha$ for $L \in [L_{\rm min}, L_{\rm max}]$, and $k_{\rm norm} = \int_{L_{\rm min}}^{L_{\rm max}} L^\alpha dL$ is its normalization factor.

Our model allows, for a fixed set of free parameters, the computation of the number of sources detected in a galaxy $N_{\rm det}$, by sampling a Poissonian with mean given by Eqn.~\ref{nobs}. The luminosities of these sources are computed numerically by sampling the XLF $N_{\rm det}$ times, with the restriction that $L > L_{\rm th}$. The sum of the sampled luminosities is $L_{\rm det}$, the total luminosity of the sources detected in the galaxy. In this way, we can compute a realization of the observables $(N_{\rm det}, L_{\rm det})$ for any galaxy with known SFR and $12 + \log ({\rm O}/{\rm H})$. We stress that our model is stochastic, i.e., each realization of our model gives a different set $(N_{\rm det}, L_{\rm det})$. Therefore, the comparison with observations must be made with care. We also stress that it is possible for our model to predict that no sources will be detectable in a given galaxy ($N_{\rm det} = L_{\rm det} = 0$), hence upper limits in the observed luminosity of galaxies can be included 
in the comparison.

\subsection{Model fitting}

The stochasticity of our model prevents us to use classical statistical methods such as least-squares fitting to compare the model to observations and fix the values of the free parameters; instead, we use a Bayesian approach. We note that the free parameter of the model may be $Q$ itself, or any set introduced to parameterize $Q$ as a function of metallicity (e.g., the slope and intercept of a linear function). We will refer to them collectivelly as $\theta$. From Bayes' theorem, the posterior probability density $f_{\rm post}$ of the model parameters $\theta$, given the observed data ($\{N, L_{\rm X}^{\rm G}\}_i$ with $i \in [1, 49]$) is

\begin{equation}
f_{\rm post}(\theta | \{N, L_{\rm X}^{\rm G}\}) = \frac{\mathcal{L}(\{N, L_{\rm X}^{\rm G}\}| \theta) f_{\rm pri}(\theta)}{\int \mathcal{L}(\{N, L_{\rm X}^{\rm G}\}| \theta) f_{\rm pri}(\theta) \, d\theta},
\end{equation}

\noindent
where $f_{\rm pri}$ is the prior probability density of the parameters, and $\mathcal{L}$ the likelihood of the data given the values of the parameters. The best-fit parameters are those that maximize the posterior probability density $f_{\rm post}$.

Assuming that we have no previous information on the parameters, we adopt a uniform prior. On the other hand, the likelihood is the product of the individual joint probability distributions of the number and luminosities of HMXBs of each galaxy,

\begin{equation}
\mathcal{L}(\{N, L^{\rm G}_{\rm X}\}) | \theta) = \prod_{i = 1}^{49} f_L(L^{\rm G}_{{\rm X},i} | N_i, \theta) P_N(N_i | \theta),
\end{equation}

\noindent
where $f_L$ is the conditional probability density for the observed luminosity given the observed number of sources, $P_N$ the probability of the observed number of sources, and the product arises because the data of different galaxies are independent. As described in the previous section,
Te best fit gives in this case the values 
\begin{equation}
P_N(N_i | \theta) = \frac{{\rm e}^{-\langle N_{\rm det} \rangle} \langle N_{\rm det} \rangle^{N_i}}{N_i!}.
\end{equation}

\noindent
Assuming that the measurement uncertainty for the luminosity $L^{\rm G}_{{\rm X},i}$ of galaxy $i$ is Gaussian with standard deviation $\sigma_{L,i}$,

\begin{equation}
\label{lumlike}
f_L(L^{\rm G}_{{\rm X},i} | N_i) = \frac{1}{\sqrt{2 \pi} \sigma_{L,i}}\int_0^\infty \exp \left( -\frac{(L^{\rm G}_{{\rm X},i} - L_{\rm det})^2}{2 \sigma_{L,i}^2}\right) g_L(L_{\rm det} | N_i) dL_{\rm det}.
\end{equation}

\noindent
Here $g_L(L_{\rm det} | N_i)$ is the probability density of the luminosity of the detected HMXBs. This factor has no analytical expression, but the integral in Eqn.~\ref{lumlike} can be computed numerically by a Monte Carlo method as the mean of the Gaussian in the integrand over a large number $M \gg 1$ of realizations of our model, i.e.,

\begin{equation}
\label{lumlikeint}
f_L(L^{\rm G}_{{\rm X},i} | N_i) = \frac{1}{\sqrt{2 \pi} \sigma_{L,i} M}\sum_{j=1}^{M} \exp \left( -\frac{(L^{\rm G}_{{\rm X},i} - L_{{\rm obs},j})^2}{2 \sigma_{L,i}^2}\right).
\end{equation}

\noindent
In this way, the likelihood $\mathcal{L}$ and hence the posterior probability density $f_{\rm post}$ can be computed within our model. The maximization of the latter was performed by a Markov-chain Monte Carlo scheme, giving the best-fit parameters and their uncertainties.

\end{document}